\documentclass[twocolumn,aps,showpacs,superscriptaddress]{revtex4}
\usepackage{mathrsfs}
\usepackage{comment}
\usepackage{amssymb}
\usepackage{amsmath}
\usepackage{amsthm}
\usepackage[dvipsnames,usenames]{color}
\usepackage{graphicx,graphics}
\usepackage{hyperref}
\usepackage{slashbox}

\begin{document}
\title{On frequency errors of nanomechanical-resonators-based-on quantum computing}

\author{Li-gong Zhou}
\affiliation{State Key Laboratory of Low Dimensional Quantum Physics£¬
Department of Physics, Tsinghua University, Beijing 100084, China
}

\author{Ming Gao}
\affiliation{State Key Laboratory of Low Dimensional Quantum Physics£¬
Department of Physics, Tsinghua University, Beijing 100084, China
}
\affiliation{Department of Physics, National
University of Defense Technology, Changsha 410073, People's Republic of China}

\author{Jin-Lin Peng}
\affiliation{State Key Laboratory of Low Dimensional Quantum Physics£¬
Department of Physics, Tsinghua University, Beijing 100084, China
}
\affiliation{School of Electronic and Electrical Engineering
University of Leeds,
Leeds,UK
}
\author{Xiang-bin Wang}
\email{xbwang@mail.tsinghua.edu.cn }
\affiliation{State Key Laboratory of Low Dimensional Quantum Physics£¬
Department of Physics, Tsinghua University, Beijing 100084, China
}
\date{\today}

\begin{abstract}
We study the consequence of the frequency errors of individual oscillators on the scalability of quantum computing based on
nanomechanical resonators. We show the fidelity change of the quantum
operation due to the frequency shifts numerically. We present a method to
perfectly compensate for these negative effects. Our method is robust to
whatever large frequency errors.


\end{abstract}

\maketitle

\hypersetup{xetex,colorlinks,linkcolor=blue,urlcolor=gray,citecolor=red,hyperindex}
\section{Introduction}
Scalability is one of the most important issues in
the realization of quantum computing.
For example, in order to factorize a 200-digit number using a quantum computer, one needs
to manipulate thousands of qubits
\cite{Cirac-Nature-2000}. However, in realistic physical systems, there are many imperfections,
such as the quantum decoherence, the device errors, and so on. These imperfections seriously limit the power of
a quantum computing device. In particular, in a large scale quantum computing, the small error of each individual
device can accumulate and this may lead to the failure of the final result \cite{Steane-2003-Overhead}.

In the recent years, methods for scalable quantum computing based on artificial quantum systems
have been extensively studied
\cite{Loss-PRA-1998,Cirac-Nature-2000,Li-PRA-2002,Hao-PRA-2007,Rabl-NaturePhys-2010}.
A promising scalable quantum computing architecture based on
spin system of nonomechanical resonators (NAMRs) was proposed by P. Rabl
\emph{et. al.} \cite{Rabl-NaturePhys-2010}. The spins have a long decoherence time and the NAMRs
can be fabricated on a large scale. The quantum motions of the NAMRs can
strongly interact with the spins \cite{PRabl-CouplingNAMRSpin-2009}
and induce strong couplings between the spins
\cite{Rabl-NaturePhys-2010,JFDu-DickeStates-2009,Zhou-NAMRSpins-PRA-2010}.

However, the inevitable fact is that the frequencies of the NAMRs can't be exactly the same in practice.
There are always frequency errors due to imperfect fabrications. Experimental results show that there can be
$\pm1.0\%$ deviations from the averaged frequency \cite{Buks-CoupledNAMR-2002}. These quantum
imperfections may cause exponential suppression of quantum
computations \cite{Keating-QuantumWalk-PRA-2007}, thus it is
an important issue  to find out quantitatively the impacts of these frequency errors and
the efficient method to compensate for the negative effects due to the device errors.

In this article, we consider a scalable quantum computing architecture consisting of $N$
spin qubits whose interactions are mediated by an array of $N$ NAMRs
\cite{JFDu-DickeStates-2009,Zhou-NAMRSpins-PRA-2010}.
The frequency errors of the NAMRs shift the frequencies of the collective modes, the
NAMR-spin coupling strengths and result in fluctuations in the
spin-spin couplings. For a given evolution time, the spins' final
state depends on the coupling strengths among them. Supposing a certain duration
is needed as the right evolution time to produce
a certain state in the ideal case
of no frequency errors, the same evolution will produce a wrong state in the actual case
with frequency errors, due to the fluctuations of the
spin-spin couplings which would reduce the quality of the quantum computing
in the spin-spin interactions mediated by the NAMRs.
Here, we analyze the impacts of the frequency errors on
the quantum operation fidelity and figure out a limitation of the
scalability of the quantum computing architecture. Our numerical
simulations show that the quantum operation fidelity decreases
rapidly with the number of the NAMR-spin elements in the present of frequency errors and these
frequency errors finally limit the scale of the system.
Based on the analysis, we propose a method to compensate for
the negative effects of whatever large frequency differences.

This article is arranged as follows: In Sec. \ref{SecModelAndMethod}
the model is given first and then the method of diagonalizing the
quadratic boson Hamiltonian is briefly reviewed. Next, in
Sec. \ref{SecOperationFidelity}, we show the
consequence of frequency errors on quantum operation fidelity by numerical simulation.
We study how the influence of the frequency errors
change with the number of the NAMRs $N$.
In Sec. \ref{SecSolution} we present our theoretical
method to resist the frequency errors.
Finally, we present discussions and conclusions  in Sec.
\ref{SecDiscussion}.

\section{Model and Method} \label{SecModelAndMethod}

\subsection{Model}
\begin{figure}
\begin{center}
  \includegraphics[width=240pt]{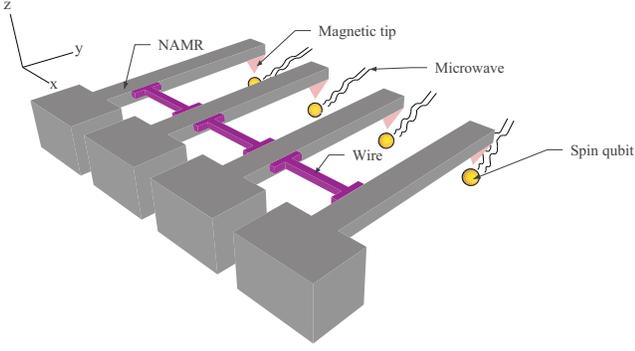}\\
  \caption{(Color online.) An NAMRs-based spin chain. The NAMRs vibrate in the $z-$direction. Each two nearest-neighbour
   NAMRs are connected by an individually isolated wire and interacted capacitively with each other.
   Each NAMR is attached with a magnetic tip on the free end and magnetically couples to a spin qubit (N-V center)
   located in the substrate below, which is driven by a local microwave.
  }\label{Fig_Model}
\end{center}
\end{figure}

As shown in Fig. \ref{Fig_Model}, we consider a system consisting of an array of $N$ NAMRs,
which are charged and interact capacitively with nearby wires interconnecting them
\cite{Rabl-NaturePhys-2010}.
A magnetic tip is attached on the free end
of each NAMR. An NAMR with fundamental frequency $\omega_i$ and
effective mass $m$ magnetically couples to an electronic spin qubit
associated with a nitrogen-vacancy (N-V) center located in the substrate below
\cite{PRabl-CouplingNAMRSpin-2009, Wrachtrup-NVCenter-2006}. Each spin is driven by a local
microwave to form a pair of dressed states in order to match the NAMR frequency
\cite{PRabl-CouplingNAMRSpin-2009}. The interaction Hamiltonian
between the NAMR and the corresponding spin qubit is $H_{\rm
sr}^{i}=\frac{\lambda}{2}(a_{i}^{\dag}+a_{i})\sigma_{z}^{i}$
\cite{Rabl-NaturePhys-2010}. Here $a_i^{\dag}$ $(a_i)$ is the
natural creation (annihilation) operator of the fundamental
vibrational mode of the $i^{\rm th}$ NAMR. $\sigma_z^i$ is the Pauli-$z$ operator
of the $i^{\rm th}$ spin. The coupling strength $\lambda=g_{\rm s}\mu_{\rm
B}G_{\rm m}a_{0}/\hbar$ with $g_{\rm s}=2$, the Bohr magneton
$\mu_{\rm B}$, the magnetic field gradient $G_{\rm m}$ and the
amplitude of zero-point fluctuations
$a_{0}=\sqrt{\hbar/(2m\omega_i)}$.

The Hamiltonian of these $N$ coupled NAMRs is
(setting $\hbar=1$) \cite{Rabl-NaturePhys-2010}:
\begin{equation}\label{Eq_Hph}
    \begin{split}
        H_{\rm ph}= &\sum_{i=1}^{N}\omega_ia_{i}^{\dag}a_{i}+\frac{1}{2}\sum_{i,j}^{N}g_{i,j}(a_{i}+a_{i}^{\dag})(a_{j}+a_{j}^{\dag}),
    \end{split}
\end{equation}
with $g_{i,j}=a_0^2[\partial^2U_{\rm el}/(\partial z_i\partial
z_j)]\vert_{z_i=0,z_j=0}$. Here $U_{\rm el}=-(U_{\rm
v}^2/2)C_{\rm \Sigma}C_{\rm w}/(C_{\rm \Sigma}+C_{\rm w})$ is
the electrostatic energy between two nearest-neighbour NAMRs
connected by a wire of self-capacitance $C_{\rm w}$ with
$C_{\rm \Sigma}=C_i(z_i)+C_j(z_j)$, $C_i(z_i)\approx C(1-z_i/h)$.
$U_{\rm v}$ is the applied gate voltage on each NAMR, $h$ is
the electrode spacing, $z_i$ is the tip position of the
NAMR and $C$ is a constant.
The first item of $H_{\rm ph}$ is the free Hamiltonian of the NAMRs
and the second item is the interactions between them.
The interaction $g_{i,j}$ consists of two parts: the
self-coupling $g_{i,i}$ and the coupling between different NAMRs
$g_{i,j}$ ($i\neq j$).

If the Hamiltonian $H_{\rm ph}$ is diagonalized by defining collective modes,
the total NAMR-spin coupling Hamiltonian (the sum of $H_{\rm sr}^i$
over $i$) can be rewritten as the coupling between the
$z$-components of the electronic spins and the collective modes of
the NAMRs. After a transformation, the effective spin-spin
interactions mediated by the NAMRs can be obtained and have the form
$H_{\rm eff}=\sum_{i,j}M_{i,j}\sigma_{z}^i\sigma_{z}^j$ with $M_{i,j}$
the coupling strength between the spin $i$ and the spin $j$, which could be
used for scalable quantum computation \cite{Rabl-NaturePhys-2010}.

\subsection{Diagonalizing the quadratic boson Hamiltonian} \label{SecDiagonzlize}
When the frequencies of the NAMRs are exactly the
same (denoted by $\omega_{\rm r}$) and the nearest-neighbour NAMR
coupling is a constant $g$, the frequencies of collective modes can be
obtained by solving the eigenvalue equations analytically and are
given by $\widetilde{\omega}_n=\sqrt{\omega_{\rm r}^2+4\omega_{\rm
r}g\{1+\cos[(n+1)\pi]/N\}}$. In practice, the
frequencies of all the NAMRs can't be exactly the same due to imperfect fabrications.
These systematic frequency errors of
each NAMR are independent and are not the same, the frequencies of
the collective modes of $N$ coupled NAMRs $\widetilde{\omega}_n$
can't be simply calculated by the perturbation method. A useful
method is described as follows.

The Hamiltonian (\ref{Eq_Hph}) can be
rewritten as
\begin{equation}
    \begin{split}
        H_{\rm ph}=\hat{\textbf{a}}^{\dag}\mathscr{D}\hat{\textbf{a}}=\sum_{k=1}^N\widetilde{\omega}_kb_k^{\dag}b_k,
    \end{split}
\end{equation}
with
$b_k^{\dag}$ $(b_k)$ the creation (annihilation)
operators for the phonons of the collective modes and
$\hat{\textbf{a}}\equiv[a_1,\cdots, a_N,
a_1^{\dag},\cdots,a_N^{\dag}]^{T}$ the natural creation
(annihilation) operators of all the NAMRs.
Here $\mathscr{D}=\left[\begin{array}{cc}A&B\\B&A\\
\end{array}\right]$, $A$ is an $N\times N$ diagonal real matrix
depending on the free Hamiltonian of these NAMRs and $B$ is an
$N\times N$ symmetric real matrix governed by the coupling model
between NAMRs.

\begin{figure}
\begin{center}
  \includegraphics[width=240pt]{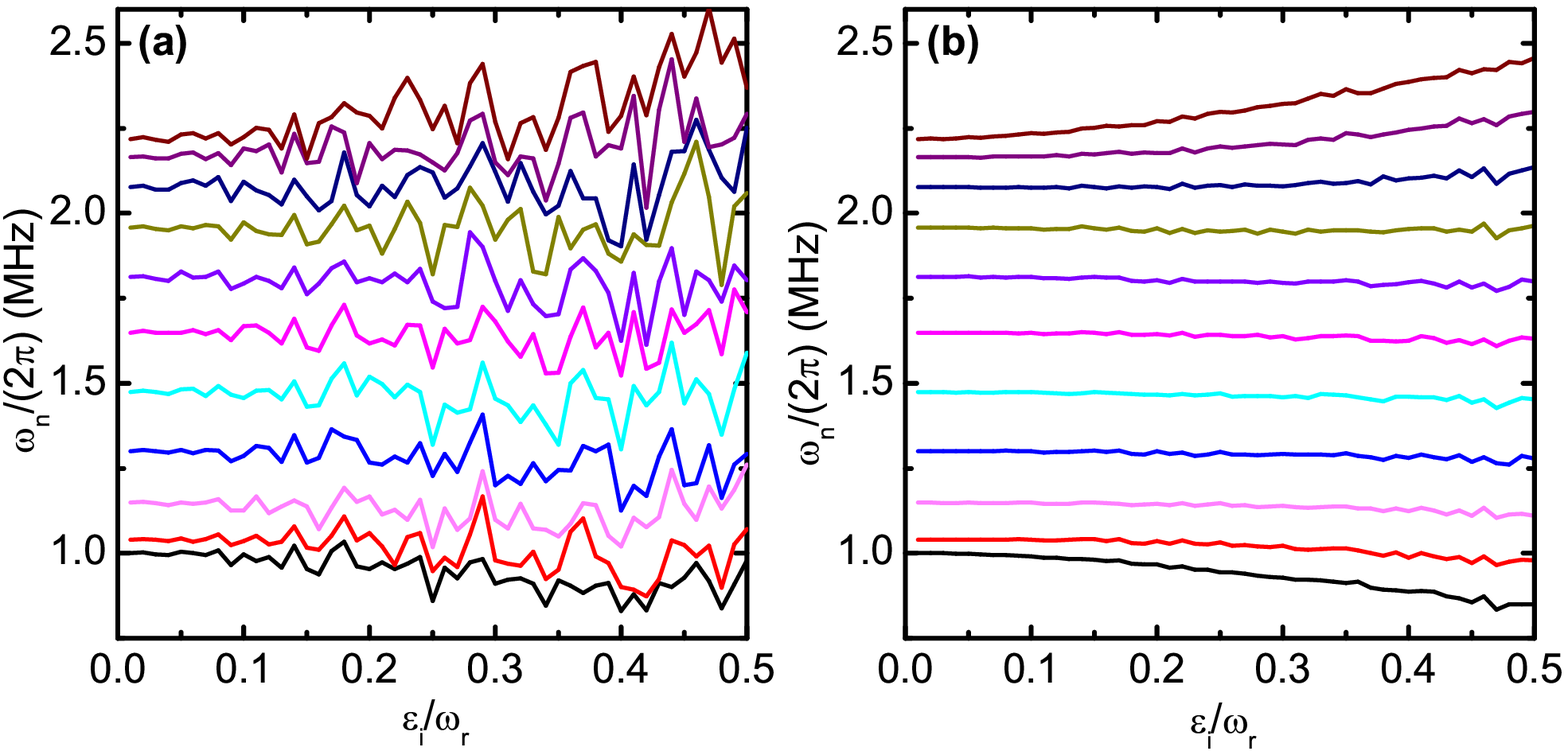}\\
  \caption{(Color online.) The collective frequencies of the NAMRs are shifted by the random frequency errors for $N=11$
  with numerical calculation. The parameters used here are: $\omega_{\rm r}/(2\pi)=1$ MHz, the coupling strength:
  $g/(2\pi)=500$ kHz. The left panel (a) is the result by running the program one time while the right panel (b)
  is the average by running the program 200 times. The offsets of the collective frequencies are optimistic
  estimations.
  }\label{Fig_wn_fr}
\end{center}
\end{figure}

Introducing an auxiliary matrix $D=A^2-B^2$, the para-values and
para-vectors \cite{Para-values} (and also the diagonalizing
para-unitary matrix \cite{Para-unitary-matrix}) of $\mathscr{D}$ can
be constructed from the eigenvalues (denoted by $d_k$ with
$k=1\cdots N$) and eigenvectors (denoted by $\xi_k$) of $D$
\cite{Colpa-DiagonalizationHamiltonian-1978}. Then the $2N\times 2N$
transformation matrix is given by $J^{-1}\equiv
[\zeta_1\zeta_2\cdots\zeta_N\zeta_{N+1}\cdots\zeta_{2N}]$ with
$\zeta_k$ and $\zeta_{N+k}$ \cite{Zeta} are $2N$-column vectors
\cite{Colpa-DiagonalizationHamiltonian-1978}. The transformation
matrix should satisfy $(J^{\dag})^{-1}\mathscr{D}J^{-1}={\rm
diag}(d_1^{1/2},d_2^{1/2},\cdots,d_N^{1/2},d_1^{1/2},d_2^{1/2},\cdots,d_N^{1/2})$
to preserve the commutation relations of bosonic operators.  The collective frequencies of these NAMRs are
\begin{equation}
    \widetilde{\omega}_k=2d_k^{1/2}.
\end{equation}
By denoting
$\hat{\textbf{b}}=[b_1,\cdots,b_N,b_1^{\dag},\cdots,b_N^{\dag}]^T $,
we obtain the relation $\hat{\textbf{a}}=J^{-1}\hat{\textbf{b}}$.
The collective frequencies are shifted by the frequency errors
and the numerically result for $N=11$ is as shown in Fig. \ref{Fig_wn_fr}.

For the system we considered,
we only need to consider the nearest-neighbour NAMR-NAMR interactions.
 The coupling strengths between nearest-neighbour
NAMRs are assumed to be the same and denoted by $g_{i,i+1}\equiv g$. The matrix $A$
and $B$ are given by \addtocounter{equation}{1}
\begin{align}
    &A=B+{\rm diag}\left(\frac{\omega_1}{2},\frac{\omega_2}{2},\cdots,\frac{\omega_N}{2}\right), \tag{\theequation a}\\
    &B=g\left[
            \begin{array}{ccccc}
                \frac{1}{2}&\frac{1}{2}&&&\\
                \frac{1}{2}&1&\frac{1}{2}&&\\
                &\ddots&\ddots&\ddots&\\
                &&\frac{1}{2}&1&\frac{1}{2}\\
                &&&\frac{1}{2}&\frac{1}{2}\\
            \end{array}
        \right]_{N\times N}.\tag{\theequation b} \label{Eq_Matrix}
\end{align}

\section{Fidelity analysis}\label{SecOperationFidelity}
Based on the model and method above, we analysize the effects of the
frequency errors. We shall calculate the fidelity of the {\em assumed state} from the ideal model without
any frequency error and the
{\em actual state} from the more realistic model with independent frequency errors in each individual NAMR.

Denoting the frequency of each NAMR $\omega_i=\omega_{\rm r}+\epsilon_i$ with $\omega_{\rm r}$
the averaged frequencies of all the NAMRs and $\epsilon_i$
the frequency error, the
coupling strength between the NAMR and the corresponding spin can be
written as
$
    \lambda_i=\lambda\sqrt{\omega_{\rm r}/\omega_{i}},
$
which means that the frequency errors generally fluctuate the
NAMR-spin couplings. The NAMR-spin coupling can be rewritten as
\begin{equation}
    \sum_i^N\frac{\lambda_i}{2}(a_i+a_i^{\dag})\sigma_z^i=\sum_{n,i=1}^N\lambda_{n,i}(b_n+b_n^{\dag})\sigma_z^i,
\end{equation}
with
\begin{equation}
    \lambda_{n,i}=\lambda_{i}(J_{i,n}^{-1}+J_{i,n+N}^{-1}),
\end{equation}
the coupling strength between the $n^{\rm th}$ collective mode and the
$z$ component of the $i^{\rm th}$ spin. Here $\{J_{i,j}^{-1}\}$ are
the matrix elements of $J^{-1}$. Therefore the effective spin-spin
coupling is
$M_{i,j}=\sum_{n}\lambda_{n,i}\lambda_{n,j}/(4\tilde{\omega}_n)$.
For a given evolution time $t_{\rm g}$, the spin-entangling
operation can be described by $U_{\rm g}(t_{\rm
g})=\exp\{i(\sum_{i}M_{i,j}\sigma_z^i\sigma_z^{j})t_{\rm g}\}$
\cite{Rabl-NaturePhys-2010}.

For $N$ NAMRs with an initial state $\vert \psi_{\rm
in}\rangle=[(\vert 0\rangle+\vert 1\rangle)/\sqrt{2}]^{\otimes N}$,
the relative frequency errors are denoted by $\Delta_i=\varepsilon_i/\omega_{\rm r}$.
After a given evolution time $t_{\rm g}$, the spins'
sate becomes $\vert \phi_{\rm o}(N, \{\Delta_i\})\rangle=U_{\rm
g}(t_{\rm g})\vert \psi_{\rm in}\rangle$. The fidelity between the {\em actual state} and the {\em ideal state} is
\begin{equation}
    F(N, \{\Delta_i\})=\vert\langle\phi_{\rm o}(N, \{\Delta_i\})\vert\phi_{\rm o}(N, \{\Delta_i=0\})\rangle\vert.
\end{equation}

\begin{figure}
  \includegraphics[width=240pt]{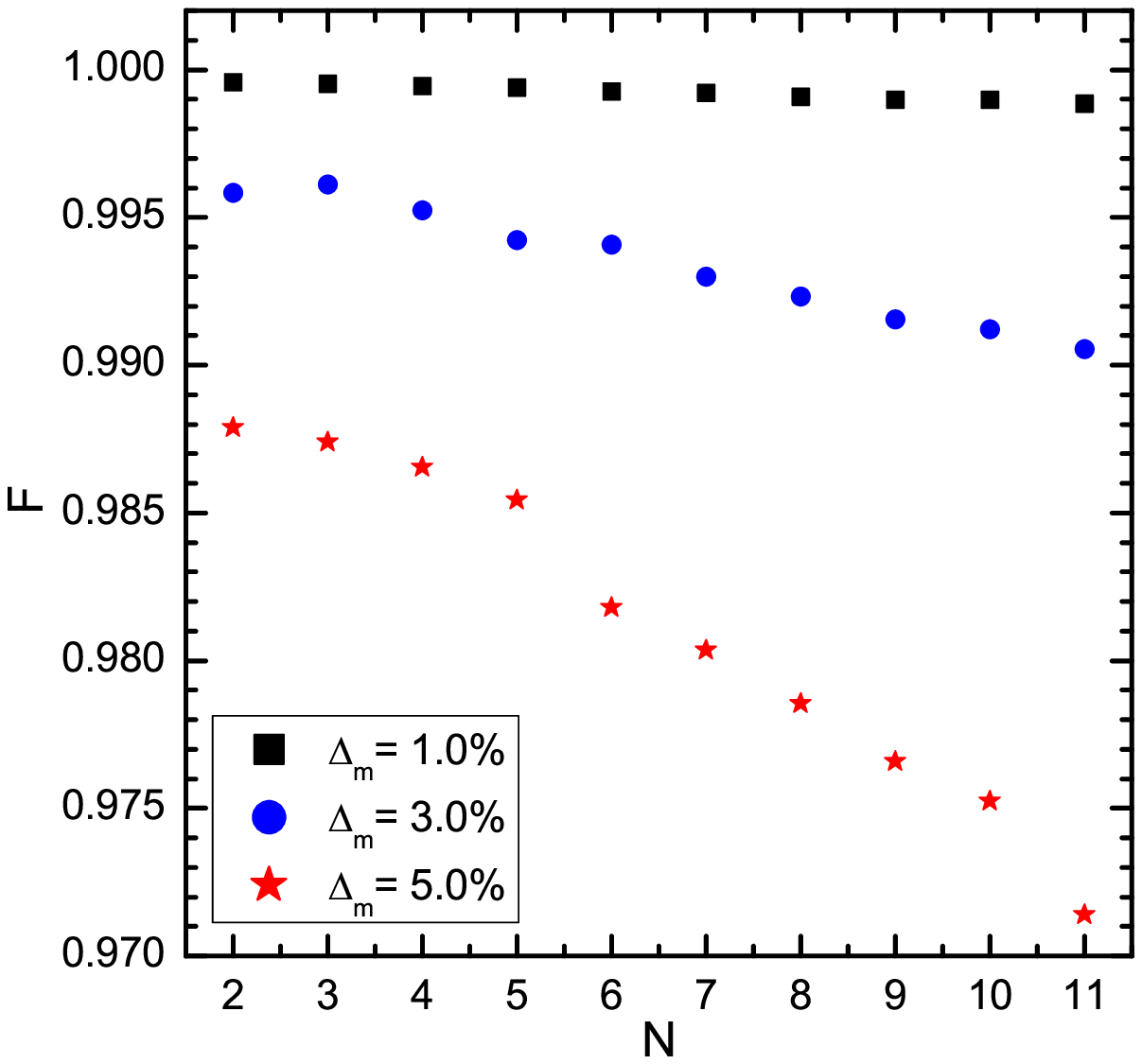}\\
  \caption{(Color online.) Fidelity decreases as the number of the NAMRs increases by numerical simulation
  (we run the simulation program $200$ times for average to get the tendency of the fidelity with the NAMR number).
  The memory and time needed for simulating quantum systems exponentially increase as the scale of the quantum system
  $N$ increases. We only simulate this system from $N=2$ to $N=11$ on an ordinary personal computer.
  The frequency errors compared with the central frequency are: 1.0\% (black solid square),
  i.e., $\Delta_{\rm m}=1.0\%$ and the same below, 3.0\% (blue solid circle), and 5.0\% (red solid star).
  }\label{Fig_FvsN_average}
\end{figure}

We consider an array of Si-NAMRs with an averaged frequency $\omega_{\rm r}/(2\pi)=1$ MHz,
$a_0\approx 1.86\times 10^{-13}$ m and $g/(2\pi)=500$ kHz
\cite{Rabl-NaturePhys-2010}. For gradient $G_{\rm m}=9.6\times
10^{6}$ T${\rm m}^{-1}$, the resulting coupling strength between an NAMR with frequency $\omega_{\rm r}$ and
a spin is about
$\lambda/(2\pi)\approx 50$ kHz. In this case, the effective
nearest-neighbour spin-spin coupling is about a few kilohertz. We
choose the evolution time $t_{\rm g}=0.3$ ms to complete one quantum
operation. (For example, given $N=2$, if we want obtain the entangled state
$(\vert 00\rangle+\vert 11\rangle+i\vert 01\rangle+i\vert
10\rangle)/2$ from the initial state $[(\vert 0\rangle+\vert
1\rangle)/\sqrt{2}]^{\otimes 2}$, the evolution time and the two
spin coupling strength $M$ should satisfy $t_{\rm g}=\pi/4\vert
M\vert$ \cite{Rabl-NaturePhys-2010}. The spin decoherence time
$T_2\approx 6$ ms \cite{Neumann-Science-6ms-2008} and the dephasing
time induced by nuclear-spin fluctuations $T_2'\approx 0.35$ ms for
N-V centers \cite{Gaebel-NatPhys-035ms-2006} are observed experimentally. )

Our numerical simulation shows that the fidelity $F(N, \{\Delta_i\})$
decreases with the number of the NAMRs $N$,  as sketched
in Fig. \ref{Fig_FvsN_average}. In Fig. \ref{Fig_FvsN_average}, the red solid star
gives $F(2,\Delta_{\rm m}=5.0\%)\approx0.988$ while
$F(11, \Delta_{\rm m}=5.0\%)\approx0.971$. $F(2,\Delta_{\rm
m}=1.0\%)\approx0.9996$ while $F(2, \Delta_{\rm
m}=5.0\%)\approx0.9879$. Here $F(N,\Delta_{\rm m}=5.0\%)$ is the value for $F(N,\{\Delta_i\})$ where the values of
each $\Delta_i$ are randomly chosen from the range $[-\Delta_{\rm m},+\Delta_{\rm m}]$.
This shows that, as the number of qubits increases,
the errors in the target state rise, hence one may end up with a wrong result with large probability
in a large scale quantum computation, even though the frequency errors of each individual NAMR are small.

\begin{figure}
  \includegraphics[width=240pt]{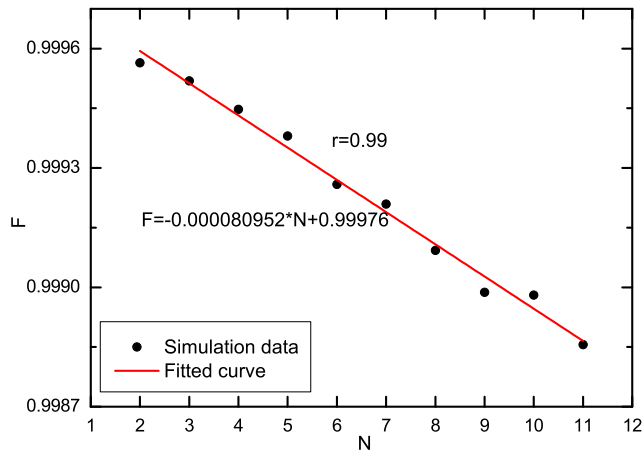}\\
  \caption{(Color online.) Fidelity $F$ is approximately linear with the number of the NAMRs $N$.
  The frequency errors here satisfy $\varepsilon_i/\omega_{\rm r}\in [-1.0\%,1.0\%]$.
  The black dots are the simulation data and the red line is the fitted curve.
  }\label{Fig_Fit_F_N}
\end{figure}

Fig. \ref{Fig_FvsN_average} also shows that $F$ is approximately linear with $N$.
For example, suppose $\Delta_{\rm m}=1.0\%$.  For $N=2\cdots 11$, the linear fitting gives the relationship
between $F$ and $N$: $F= -0.000080952\times N+ 0.99976$ with the
correlation coefficient $r=0.99$ as shown in Fig. \ref{Fig_Fit_F_N}.
For a given $F_0$, if we want the fidelity $F>F_0$, the length of a spin chain should
be less than about $\lfloor
10^{4}\times(0.99976-F_0)/0.80952\rfloor$ spins. The symbol $\lfloor
P \rfloor$ denotes the maximal integer not great than $P$. For
example, we obtain $N<4.6\times10^3$ for $F_0=2/3$. It can be seen
that the frequency errors do limit the maximum length of the spin
chain if we want to produce a faithful resultant state and finally limit the extensibility of
the fault-tolerant quantum computing.

\section{Compensation method}\label{SecSolution}
\begin{center}
\begin{figure}
  \includegraphics[width=180pt]{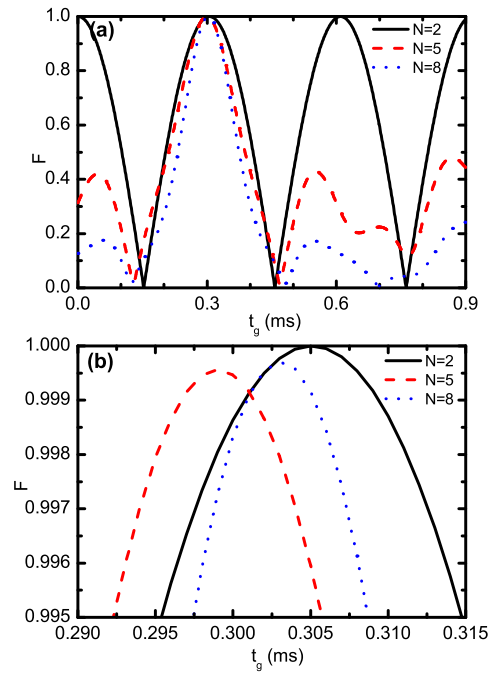}\\
  \caption{(Color online.) The fidelity $F$ changes as a function of the evolution time $t_{\rm g}$
  in the case of frequency errors for a fixed $N$. We set the system state at the time $t_{\rm g}=t_{\rm g0}=0.3$ ms
  without frequency errors as the ideal final state. $F$ denotes the fidelity between the ideal final state and
  the actual state at an arbitrary time $t_g$. The infidelity $(1-F)$ caused by frequency errors can't be compensated for
  by only adjusting the evolution time $t_{\rm g}$ when $N>2$. In this figure, the frequency errors are
  within $1.0\%$ compared with $\omega_{\rm r}$. The black (solid) curve is for $N=2$, the dashed (red) curve
  is for $N=5$ and the dotted (blue) curve is for $N=8$. The curves in panel (b) are the plot details of panel (a)
  for the regime $t_{\rm g}\in [ 0.290,0.315]$ ms.}\label{Fig_F_t}
\end{figure}
\end{center}

As analyzed above, the frequency errors can lead to
fidelity decease of the quantum state evolution. The deviation cannot
be compensated by simply adjusting the evolution time $t_{\rm g}$ when
$N>2$, as shown in Fig. \ref{Fig_F_t}. In what follows, we propose a
method to solve this problem. The main idea is that the errors caused by
frequency differences can be perfectly compensated for by controlling
the interaction time of each two adjacent spin qubits block. With this compensation method, one can produce
arbitrarily large-scale states with whatever large frequency error of each NAMR.

\begin{figure}
\begin{center}
  \includegraphics[width=140pt]{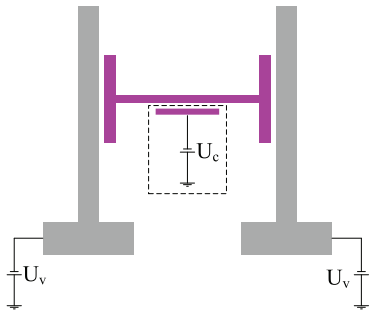}\\
  \caption{(Color online.) Switchable coupling between two NAMRs (top view). A switchable control gate (dashed box)
  is added between every two nearest-neighbour NAMRs. For a control voltage $U_{\rm c}=0$, the two NAMRs couple
  with each other. For $U_{\rm c}=U_{\rm v}$, there is no energy associated with charge flowing from
  the gate capacitors onto the wire and thereby the coupling between the two NAMRs is switched off.
  }\label{Fig_SwitchableGate}
\end{center}
\end{figure}

For a one-dimensional $N$ NAMR-spin chain, an auxiliary switchable
control voltage gate is added between every two nearest-neighbour
NAMRs to obtain switchable coupling between them, as depicted in Fig. \ref{Fig_SwitchableGate}. When the control
voltage is zero, the nearest-neighbour NAMRs couple with each other
and induce effective spin-spin interactions; and when the control
voltage is switched on and equals to the NAMR's gate voltage, the
coupling of the two NAMRs is switched off, hence the NAMRs induce no effective spin-spin
interactions. In addition, we assume that we have obtained the
fundamental frequencies of the NAMRs by average after multiple
measurements (the systematic errors still exist due to the
limitations of the fabrication techniques and the random measurement
frequency errors are eliminated by average).
The effective coupling strength
between every two nearest-neighbour spins in the case of frequency
errors can be theoretically calculated by the formula $M_{i,i+1}$
by assuming that the other NAMRs and spins do not exist.

\begin{figure}
\begin{center}
  \includegraphics[width=200pt]{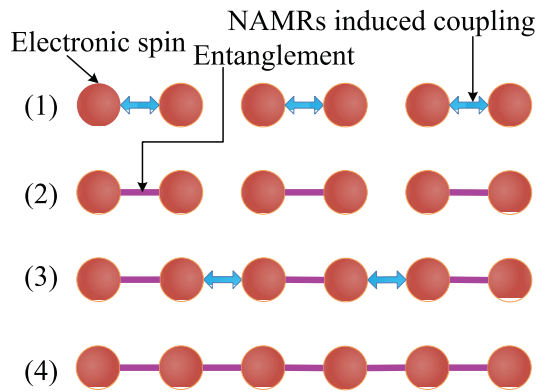}\\
  \caption{(Color online.) A sketch for the procedures to compensate for the frequency errors.
  }\label{Fig_Solution}
\end{center}
\end{figure}

For simplicity, we assume $N$ is an even number (for an odd $N$,
we only need to regard the last spin as an already entangled spin pair).
One can achieve the perfect result through the following procedures as sketched in Fig. \ref{Fig_Solution}:
(1) switch on the coupling between the $(2i-1)^{\rm th}$ and
the $(2i)^{\rm th}$ spins for $i=1\cdots N/2$ and switch off the coupling
between the $(2i)^{\rm th}$ and the $(2i+1)^{\rm th}$ spins for $i=1\cdots (N/2-1)$.
(2) control the evolution  times between each two-spin pair to obtain
the ideal entanglement between them and then switch off the coupling, respectively.
(3) switch on the coupling between the $(2i)^{\rm th}$ and the $(2i+1)^{\rm th}$ spins
for $i=1\cdots (N/2-1)$ when all the couplings are switched off in (2).
(4) repeat (2). The total time needed for all these
processes is approximately two times of that in the case of no frequency errors.
In addition, this method also avoid the influence of effective sub nearest-neighbour interactions between the spins.

\begin{proof}
The operators of different spins communicate with each other. In the
case of no frequency errors, the NAMR-spin couplings reduce to a
constant and denoted by $\lambda$. All the effective nearest-neighbour spin-spin couplings
induced by the NAMRs are equal to each other and denoted by $M$. According to the Ising model, the
evolution operator of the spin chain for a given evolution time
$t_{\rm g}$ is:
\begin{equation}
\begin{split}
    U_{\rm g}(t_{\rm g})=&\exp\left\{i\sum_{i=1}^{N-1}M\sigma_{z}^i\sigma_z^{i+1}t_{\rm g}\right\}\\
    =&\prod_{i=1}^{N-1}\exp\left\{iM\sigma_{z}^{i}\sigma_{z}^{i+1}t_{\rm g}\right\}.\\
\end{split}
\end{equation}

In the case of frequency errors, the NAMR-spin couplings and the
collective modes of the NAMRs are shifted by the frequency errors.
Denoting the coupling strength between the $i^{\rm th}$ and
$(i+1)^{\rm th}$ spins as $M_{i,i+1}$, the ideal entanglement between
two spins can be obtained by adjusting the evolution time in the
two spin case as shown in Fig. \ref{Fig_F_t} and the time needed is
denoted by $t_{i,i+1}$. When all the four steps are completed,
the evolution operator of these spins can be described by:
\begin{equation}
\begin{split}
    U_{\rm g}(t_{1,2}\cdots t_{N-1,N})=&\prod_{i=1}^{N-1}\exp\left\{iM_{i,i+1}\sigma_{z}^{i}\sigma_z^{i+1}t_{i,i+1}\right\}.\\
    \end{split}
\end{equation}
By precisely controlling the evolution time $t_{i,i+1}=(Mt_{\rm
g})/M_{i,i+1}$, the evolution operators in the two cases are equal
to each other and the two final states must be the same if the
system evolutes from the same initial state.
\end{proof}

We should point out that the entanglement of the spins prepared by the following steps is not perfect:
(1) switch on all the nearest-neighbour NAMR-NAMR couplings. (2) switch off the coupling
when the nearest-neighbour spins involute to the maximum entanglement sates
(the effective interaction between the nearest-neighbour spins can be numerically calculated), respectively.
A spin interacts with all the other spins through the collective modes of the NAMRs when the control voltage
is zero. These interactions inevitably include sub nearest-neighbour interactions and so on.
These sub nearest-neighbour interactions decay as $M_{i, i\pm m}\sim (g/\omega_{\rm r})^{(m-1)}$
\cite{Rabl-NaturePhys-2010},
but their influences are generally comparable with those due to frequency errors.
Unwanted entanglement between the spins which are not nearest-neighbours is generated and
the Ising interaction produced by the above two steps is unsatisfactory.

\section{Discussions and conclusions}\label{SecDiscussion}
Although only the one-dimensional
case is discussed in the context, the similar analysis on the
two-dimensional NAMRs-spins quantum computing architecture
\cite{Rabl-NaturePhys-2010} can also be handled by using the same method.
In the two-dimensional configuration, NAMRs are ordered on a two-dimensional lattice and
couple to their four neighbours electrostatically.
In the two-dimensional case, the effects of the frequency errors
can be analysed by adjusting the matrix
$B$ in equation (\ref{Eq_Matrix}) to be a general matrix instead of
a tridiagonal matrix when considering the nearest-neighbour
couplings. In the compensation method, a switchable voltage gate should
be added between each two nearest-neighbour NAMRs.

In summary, we study the impacts of the frequency differences on the scalability of a promising NAMRs-based-on
quantum computing architecture. The influences on the quantum operation fidelity are analyzed in detail,
and a method is given to compensate for the
negative effects of these frequency differences.

\section*{ACKNOWLEDGMENTS}
The authors thank Wenjie Zou, Jiazhong Hu for useful discussions and valuable advices. This work was supported in part by the National Basic Research Program of China grant No. 2007CB807900 and 2007CB807901, NSFC grant No. 60725416.


\begin{thebibliography}{99}
\bibitem{Para-values}Para-values and para-vectors. Let $\mathscr{D}=\left[\begin{array}{cc}X&Y\\Z&W\\\end{array}\right]$ with
X, Y, Z, W four $N\times N$ matrices. If
$\mathscr{D}\left[u_{\rho},v_{\rho}\right]^{T}=\lambda_{\rho}\left[u_{\rho},-v_{\rho}\right]^{T}$
is stasified, where $u_{\rho}$, $v_{\rho}$ are $N$-column vectors with
$\rho=1,2,\cdots,2N$, we call $\lambda_{\rho}$ the para-values and
$\left[u_{\rho},v_{\rho}\right]^{T}$ the
para-vectors of the matrix $\mathscr{D}$.
\bibitem{Para-unitary-matrix} Para-unitary matrix. Define the para-unit matrix $\mathscr{I}\equiv{\rm diag}(1,1,\cdots,1,-1,-1,\cdots,-1)_{2N\times
2N}$. We call $\mathscr{J}$ a $2N\times 2N$ para-unitary matrix if
$\mathscr{J}^{\dag}\mathscr{I}\mathscr{J}=\mathscr{I}$.
\bibitem{Zeta} The vectors $\zeta_k$ and $\zeta_{N+k}$ are given by
$\zeta_k=C_k \left[(A+d_k^{1/2}I)\xi_k,-B^{\dag}\xi_k\right]^{T} \equiv \left[u_k,v_k\right]^{T}$,
$\zeta_{N+k}=C_{N+k} \left[-B\xi_k,(A+d_k^{1/2}I)\xi_k\right]^{T} \equiv \left[u_{N+k},v_{N+k}\right]^{T}$.
Here, $I$ is the $N\times N$ unit matrix and $C_k$, $C_{N+k}$ are determined such that the vectors are para normalized : $u_k^{\dag}u_k-v_k^{\dag}v_k=1$ and $u_{N+k}^{\dag}u_{N+k}-v_{N+k}^{\dag}v_{N+k}=1$.
\end{thebibliography}
\end{document}